\documentstyle[aps]{revtex} 
\begin{document} 
\title{Vibrational and Acoustical Properties of a Liquid 
Drop In the Phase-Separated Fluid With a Highly Mobile Interface
}
\author{S.N.Burmistrov$^{1,2}$ and T.Satoh$^1$}
\address{ $^1$Department of Physics, Faculty of Science, Tohoku 
University, Sendai 980, Japan }
\address{$^2$Theoretical Division, Kurchatov Institute, Moscow 123182, Russia
} 
\date{ \today}
\maketitle 
\begin{abstract}
We study the oscillation spectrum and acoustic properties of a liquid 
drop in the phase-separated fluid when the interfacial dynamics of 
phase conversion can be described in terms of the kinetic growth coefficient. 
For a readily mobile interface, i.e., as the growth coefficient becomes 
comparable with a reciprocal of the acoustic impedance, anomalous behavior is 
found in the oscillation spectrum of a drop as well as in the velocity and 
absorption of a sound wave propagating through a suspension of drops in the 
two-phase system. Compared with the known case of two immiscible fluids, the 
high interface mobility leads to an anomalous softening of the radial drop 
pulsations and to the frequency- and temperature-dependent behavior for the 
sound velocity and absorption coefficient in a two-phase suspension. 
\end{abstract}  
({\bf 1}) Below 0.87K a liquid $^3$He-$^4$He mixture is known to separate 
into the $^3$He-concentrated ({\it c}) and $^3$He-dilute ({\it d}) phases. 
Thus the superfluid-normal interface between the {\it c}- and {\it d}-phases of 
a phase-separated liquid mixture can represent a system for studying 
the dynamic properties of the phase-separation in a supersaturated liquid 
mixture at low temperatures down to absolute zero. Recently, the interface 
has been observed to exhibit an anomalous behavior in the transmission of 
sound at frequency $\approx$10MHz below 70mK \cite{1SS}. The sound 
transmission coefficient at $T<$10mK reduces by more than 10 times compared 
with its high temperature value corresponding to the classical 
acoustic-mismatch theory. 
\par The amplitude of a transmitted sound wave is directly determined by the 
boundary conditions for the phase conversion at the interface. For the sound 
transmission coefficient at normal incidence, 
\begin{equation}
\tau _{1\rightarrow 2} = \frac{2Y_2}{Y_{1}+Y_{2}+Y_{1}Y_{2}\xi} 
\label{equ1}
\end{equation}
where $Y_{1,2}=\rho _{1,2}c_{1,2}$ are the acoustic impedances 
given by a product of the density $\rho _{1,2}$ and sound velocity 
$c_{1,2}$ for the {\it c}- and {\it d}-phases, respectively. The physical 
meaning of the kinetic growth coefficient $\xi$, which vanishes for the two 
immiscible liquids, can be defined through the relation 
\begin{equation}
J = \xi\left(\frac{\rho _{1}\rho _{2}}{\rho _{1} -\rho _{2}}\right) ^{2}
\Delta\mu 
\label{equ2}
\end{equation}
Here $\Delta\mu =\mu _{1}-\mu _{2}$ is the difference between $\mu _1$ and 
$\mu _2$, the chemical potentials per unit mass of the {\it c}- and 
{\it d}-phases, respectively, and $J$ is the mass flow across the interface 
per unit area. Thus the anomalous reduction of the sound transmission 
evidences that the interface between the {\it c}- and {\it d}- phases of a 
mixture can be highly mobile at sufficiently low temperatures. In fact, the 
growth coefficient $\xi =\xi (T)$ increases by more than 3 orders of the 
magnitude, reaching $Y_2\xi\approx 20\gg 1$ as $T<$10mK \cite{1SS}. Here we 
consider the effect of the high mobility of the interface on (a) the spectrum 
of small oscillations of a {\it c}-drop and (b) the velocity and absorption 
of sound propagating across a suspension of {\it c}-drops in the {\it d}-phase. 
\newline
({\bf 2}) Let us turn first to the oscillation spectrum of a drop. 
The problem can be stated as follows. A fluid of 
density $\rho _1$ is contained within a sphere of radius $R$ and the other 
fluid phase of density $\rho _2$ occupies the exterior region confined by a 
rigid wall of radius $R_b\gg R$ so that the fluid cannot penetrate through the 
wall. The phase conversion rate at the interface is assumed to be governed 
by Eq. (\ref{equ2}). 
\par In the bulks of the fluids the velocity potentials $\phi _{1,2}$ 
satisfy, respectively, 
\begin{equation}
\nabla ^{2}\phi _{1,2} -\ddot{\phi} _{1,2}/c^{2}_{1,2} =0
\label{equ3}  
\end{equation} 
As usual,  the deformation of the interface $\zeta$ is supposed to be small 
and the radius of a drop is described by 
\begin{equation}
R_{\zeta}(t)=R+\zeta _{l}(t)Y_{l}
\label{equ4}
\end{equation} 
Here $Y_l$ is a spherical harmonic function of order $l=0, 1, 2, ...$  and 
we put $\phi _{1,2}(t)$ and $\zeta (t)\propto\exp (-\imath\omega t)$. 
\par The analysis will be restricted to first order in $\zeta$ and the 
pressure balance for the disturbed magnitudes of pressure $\delta P_{1,2}=
-\rho _{1,2}\dot{\phi} _{1,2}(r, t) $ at the interface reads 
\begin{equation} 
\delta P_{1} - \delta P_{2} = \alpha\frac{(l-1)(l+2)}{R^2}\zeta _{l} 
\label{equ5}
\end{equation}
where $\alpha$ is the surface tension and the other factor is a change of the 
curvature of the interface \cite{2LL}. In addition, we must involve a 
continuity of the mass flow of the fluids across the interface and according 
to (\ref{equ2}) we have 
\begin{eqnarray}
\rho _{1}(v_{1,r}-\dot{\zeta}) = \xi\left(\frac{\rho _{1}\rho _{2}}
{\rho _{1} - \rho _{2}}\right) ^{2}\, 
\left(\frac{\delta P_1}{\rho _1} - \frac{\delta P_2}{\rho _2}\right)
\nonumber\\
\rho _{2}(v_{2,r}-\dot{\zeta}) = \xi\left(\frac{\rho _{1}\rho _{2}}
{\rho _{1} - \rho _{2}}\right) ^{2}\, 
\left(\frac{\delta P_1}{\rho _1} - \frac{\delta P_2}{\rho _2}\right)
\label{equ6}
\end{eqnarray} 
Since we consider the total bulk to be closed, the above boundary conditions 
should be augmented by vanishing the fluid velocity at the wall, i.e., 
$\nabla\phi _{2}(r=R_b)=0$. 
\par Let us first turn to the case of the spherically symmetrical pulsations 
corresponding to zero harmonic $l=0$. Eliminating $\zeta$ in (\ref{equ5}) and 
(\ref{equ6}), we arrive at the following dispersion equation 
\begin{equation}
\omega ^{2}
\left(\rho _{2}\Lambda _{2} -
\rho _{1}\Lambda _{1}\right) -\frac{2\alpha}{R^2} = 
\imath\omega\rho _{1}\rho _{2}\xi 
\left[\omega ^{2}\Lambda _{1}\Lambda _{2} + 
\frac{2\alpha}{R^2}\frac{\rho _{1}\rho _{2}}{(\rho _{1}-\rho _{2})^2}
\left(\frac{\Lambda _2}{\rho _2} - \frac{\Lambda _1}{\rho _1}\right)
\right] 
\label{equ7}
\end{equation}
The functions $\Lambda _{1,2}$ are ratios of $\phi _{1,2}$ to velocities 
$v_{1,2}=\nabla\phi _{1,2}$ taken at the interface $r=R$ 
\begin{eqnarray}
\Lambda _{1} & = & \frac{R\sin q_{1}R }{q_{1}R\cos q_{1}R - \sin q_{1}R }
\nonumber \\
\Lambda _{2} & = & \frac{-R[\sin q_{2}(R_{b}-R) - 
q_{2}R_{b}\cos q_{2}(R_{b}-R)] }
{q_{2}R[\cos q_{2}(R_{b}-R) +q_{2}R_{b}\sin q_{2}(R_{b}-R)] + 
\sin q_{2}(R_{b}-R) -q_{2}R_{b}\cos q_{2}(R_{b}-R) }   
\label{equ8}
\end{eqnarray} 
where the  wave vectors are $q_{1,2}=\omega /c_{1,2}$.   
\par To clarify the essentials, let us first compare the two limiting cases, 
namely, immiscible fluids when $\xi =0$ and readily miscible ones when 
$Y_{2}\xi\gg 1$. In the first case the oscillation spectrum is well known 
\cite{2LL}. Provided the both fluids are incompressible, no radial pulsations 
are possible in the system since any change of the drop volume is strictly 
forbidden. Involving the 
compressibility results in the appearance of oscillations with the minimum 
frequency of about $\omega\approx 4.5c/R_b$ ($c_1\approx c_2\approx c$). The 
effect of the surface tension is negligible while we do not consider the 
drop radius smaller than $R_{\alpha}=2\alpha /\rho _{1}c_{1}^{2}$. For our 
case, $R_{\alpha}\approx 0.5\AA$ due to the smallness of 
$\alpha =$0.0239 erg/cm$^2$ \cite{3SS}.
\par A specific feature of the opposite case $Y_{2}\xi\gg 1$ is an emergence 
of a soft mode in the oscillation spectrum which can lead to an instability 
of sufficiently small-sized drops. In fact, for the frequencies $\omega\ll 
c_{2}/R_{b}$, Eq. (\ref{equ7}) reduces to 
\begin{equation}
\omega ^{2} + \imath\frac{\omega}{Y_{2}\xi}\frac{c_2}{R} 
+ \frac{2\alpha}{R^3}\frac{\rho _2}{(\rho _{1}-\rho _{2})^2} 
-\frac{3R}{R_b}\left(\frac{c_2}{R_b}\right) ^{2} =0 
\label{equ9}
\end{equation}
As is seen, in the $Y_{2}\xi =\infty$ limit only the drops which size exceeds 
the critical one 
\begin{equation}
R_{c}=R_{b}\left(\frac{2\alpha\rho _{2}}
{3(\rho _{1}-\rho _{2})^{2}c_{2}R_b}\right) ^{1/4}
\end{equation} 
are stable against the radial pulsations. The spectrum for drops of 
$R\gg R_c$ also is enormously softened since $\omega _{min}=
(3R/R_{b})^{1/2}c_{2}/R_b\ll c_{2}/R_b$. The reason of such softening lies in 
the appearance of a pressure node $\delta P _{1,2}=0$ at the interface 
for the infinite growth coefficient, smoothing the pressure variation in 
the bulk. 
\par The finite magnitude of the growth coefficient brings a damping into 
the drop pulsations and acts as a stabilizing factor. Provided $R>R_c$, the 
radial pulsations can be either quasiperiodic with the damping coefficient 
$(2Y_{2}\xi)^{-1}c_{2}/R$ for $2Y_{2}\xi >(R_{b}/R)^{3/2}\gg 1$ or aperiodic 
if $R_{b}/R<2Y_{2}\xi <(R_{b}/R)^{3/2}$. For small-sized drops $R<R_c$, the 
instability remains but a positive imaginary part of $\omega$ responsible 
for the instability strongly decreases. On the whole, if $Y_{2}\xi
\rightarrow 0$ and radius $R\approx R_c$, 
\begin{equation}
\omega = -\imath Y_{2}\xi\,\frac{3R_{c}^{2}}{R_{b}^{2}}
\left(\frac{R^2}{R_{c}^2}-\frac{R_{c}^2}{R^2}\right)\frac{c_2}{R_b}
\label{equ11}
\end{equation}
\par For the high frequency modes of radial pulsations at 
$\omega\sim c_{2}/R_{b}$ or $c_{1}/R$, the contribution from the surface 
tension can be neglected. The finite magnitude of the growth coefficient 
leads to an additional damping which is maximum at $Y_{2}\xi\sim 1$ and to 
a temperature-dependent shift of the pulsation frequency due to $\xi =\xi (T)$. 
\par To gain some insight, we put $c_{1}=c_2$ and $\rho _{1}=\rho _2$, 
reducing Eq.(\ref{equ7}) to 
\begin{equation}
\sin qR_{b} - qR_b\cos qR_{b} = \imath Y\xi\sin qR \, 
\left[ \sin q(R_{b}-R) - qR_b\cos q(R_{b}-R)\right] 
\label{equ11a}
\end{equation} 
In the limit $Y\xi\ll 1$ we have a positive shift in frequency 
\begin{equation}
\left(\omega -\omega _0\right)/\omega = 
- (\imath -Y\xi qR)Y\xi \left( qR_{b}+1/qR_b\right) \left( R/R_b\right) ^2 
\label{equ11b}
\end{equation}
where $qR_b$ is one of the roots of equation $\tan x = x$. In the opposite 
case $Y\xi\gg 1$, the shift of the frequency compared with $\omega _{\infty}$ 
at $Y\xi =\infty$ is negative and we obtain for the frequencies of about 
$c_{2}/R_b$ 
\begin{equation}
\frac{\omega -\omega _{\infty} }{\omega } = 
-\frac{1}{Y\xi}\left(\imath +\frac{1}{Y\xi qR}\right)
\frac{1+(qR_b)^2}{(qR_b)^3} 
\label{equ11c}
\end{equation}
and for the frequencies of about $c_{1}/R$ 
\begin{equation}
\frac{\omega -\omega _{\infty}}{\omega } = 
-\frac{1}{Y\xi qR}\left(\imath +\frac{1}{Y\xi}\,
\frac{1+qR_b\tan qR_b}{\tan qR_{b} - qR_b}\right) 
\label{equ11d}
\end{equation}
\par Concerning nonspherical pulsations $l\neq 0$, we must note that 
such type of oscillations is not accompanied by changing the volume of a drop. 
Thus we can neglect the compressibility of the both fluids and consider 
the oscillations in the infinite bulk of the {\it d}-phase with $R_b=\infty$. 
As a result, we find that nonspherical oscillations are described by an 
equation inherent in oscillatory processes with a single relaxation time 
\begin{equation}
\omega ^{2} - \omega _{0}^{2} =\imath\omega\tau (\omega ^{2} - 
\omega _{\infty}^{2}) 
\label{equ12}
\end{equation}
Here $\tau$ plays a role of some effective relaxation time for the drop 
pulsations 
\begin{equation}
\tau =\tau _{l} =\frac{\rho _{1}\rho _{2}R\xi}{(l+1)\rho _{1}+l\rho _{2}} 
\label{equ13}
\end{equation}
The frequency $\omega _{0}$ 
\begin{equation}
\omega _{0}^{2}=\frac{(l-1)l(l+1)(l+2)}{(l+1)\rho _{1} +l\rho _{2}}\cdot
\frac{\alpha}{R^3}
\label{equ14}
\end{equation}
corresponds to the low frequency $\omega\tau\ll 1$ limit and is the well-known 
oscillation frequency of a liquid drop in a medium of two immiscible fluids, 
e.g., \cite{4RA}. In the high frequency limit $\omega _{\infty} >\omega _0$ 
and  
\begin{equation}
\omega _{\infty}^{2}=(l-1)(l+2)\frac{l\rho _{1}+(l+1)\rho _{2}}
{(\rho _{1}-\rho _{2})^2}\cdot \frac{\alpha}{R^3} 
\label{equ15}
\end{equation}
\par Unlike $l=0$, the oscillations of the shape of a drop in the highly 
miscible fluids prove to be always stable. For $l=2$, 
$\omega _{\infty}/\omega _{0}\approx 3.7$ and, if $R=1\mu$m, 
$\omega _{\infty}\approx 3.9$MHz. Since the latter is very close to the 
experimental conditions in \cite{1SS}, we may expect the high frequency 
behavior for the oscillations of 1$\mu$m-sized {\it c}-phase droplets below 
about 30mK. 
\newline 
({\bf 3}) Here we shall analyze some acoustic properties of {\it c}-drops, 
namely, scattering of sound wave with a {\it c}-drop and the velocity of 
the sound propagating through a suspension of {\it c}-drops in the 
{\it d}-phase. In the last case we suppose that the scattering with each 
{\it c}-drop occurs independently from the others. We employ the analogy 
with optics used for studying acoustic properties in immiscible fluids 
\cite{5Ak}. 
\par It is convenient to introduce the acoustic refraction index 
\begin{equation}
n = 1+\frac{2\pi}{q^{2}V}\sum _{R} f_{R}(0)
\end{equation}
where a sum is taken over all {\it c}-drops, $V$ is a volume of the system, 
and $f_{R}(0)$ is the zero-angle scattering amplitude on a single drop of 
radius $R$. The relative change of the sound velocity and absorption 
coefficient are given by 
\begin{equation}
\frac{\Delta c_2}{c_2} =1 - Re\, n = -\frac{2\pi}{q^{2}V}\sum _{R}Re\, f_{R}(0)
\,\, ;\,\,\,\, \gamma =\frac{\omega}{c}Im\, n
\label{equ17}
\end{equation} 
Using the optical theorem \cite{6LL}, we estimate the sound absorption 
coefficient as 
\begin{equation}
\gamma = \frac{1}{2V}\sum _{R}\sigma _{tot}(R)
\label{equ18}
\end{equation}
where $\sigma _{tot}(R)$ is the total cross-section of scattering with a 
drop of radius $R$. 
\par Since the scattering amplitude is a sum of the partial amplitudes 
\begin{equation}
f(0) =\sum _{l=0}^{\infty} (2l+1)f_{l}(0)
\label{equ19}
\end{equation}
correspondingly, we can represent the relative change of the sound velocity 
and absorption coefficient as a sum of the partial contributions 
\begin{equation}
\Delta c_{2}/c_{2}=\sum _{l=0}^{\infty}\Delta c_{2}^{(l)}/c_{2} \,\, ;\,\,\,\, 
\gamma =\sum _{l=0}^{\infty}\gamma ^{(l)}
\label{equ20}
\end{equation}
\par To solve the scattering problem, we use the same 
Eqs. (\ref{equ3}--\ref{equ6}). The velocity potential outside the drop at 
$r>R_{\zeta}(t)$ should be represented as 
\begin{equation}
\phi _{2}(r, t)=\sum _{l=0}^{\infty} (2l+1)[B_{l}\jmath _{l}(q_{2}r) + 
C_{l}y_{l}(q_{2}r)]\exp (-\imath\omega t)
\label{equ21}
\end{equation}
where $\jmath _{l}(x)$ and $y_{l}(x)$ are the spherical Bessel functions of 
first and second kind, respectively. The partial scattering amplitude is 
determined by a ratio of $C_{l}/B_{l}$ 
\begin{equation}
f_{l} = \frac{\imath}{q_2}\,\frac{C_{l}/B_{l}}{C_{l}/B_{l} -\imath} =
\frac{\imath}{q_2}\,\frac{A_l}{A_{l} -\imath (1-\omega _{l}^{2}/\omega ^2)}
\label{equ22}
\end{equation}
where 
\begin{eqnarray}
A_{l} & = & \left( \rho _{1}q_{2}\jmath _{l}(1)y'_{l}(2) -
\rho _{2}q_{1}\jmath '_{l}y_{l}(2) +
\imath\omega\xi\rho _{1}\rho _{2}\jmath _{l}(1)y_{l}(2) \right)^{-1} 
\nonumber \\
& \times & \left\{\rho _{2}q_{1}\jmath '_{l}(1)\jmath _{l}(2) 
-\rho _{1}q_{2}\jmath _{l}(1)\jmath '_{l}(2)
- \imath\omega\xi\rho _{1}\rho _{2}\jmath _{l}(1)\jmath _{l}(2) 
 +  \frac{\alpha (l-1)(l+2)}{\omega ^{2}R^2} \right. \nonumber\\
& \times & \left. \left[ q_{1}q_{2}\jmath '_{l}(1)\jmath '_{l}(2) - 
\frac{\imath\omega\xi}{R}\left(\frac{\rho _{1}\rho _{2}}
{\rho _{1}-\rho _{2}}\right) ^{2}
\left(\frac{q_2}{\rho _1}\jmath _{l}(1)\jmath '_{l}(2) - 
\frac{q_1}{\rho _2}\jmath '_{l}(1)\jmath '_{l}(2)\right)\right]\right\} 
\label{equ23}
\end{eqnarray}
and the resonance frequency 
\begin{eqnarray}
\omega _{l}^{2} & = & \left( \alpha (l-1)(l+2)/R^2\right) \left[ 
\rho _{1}q_{2}\jmath _{l}(1)y'_{l}(2) - \rho _{2}q_{1}\jmath '_{l}(1)y_{l}(2) 
+\imath\omega\xi\rho _{1}\rho _{2}\jmath _{l}(1)y_{l}(2)\right] ^{-1} 
\nonumber\\
& \times & 
\left\{ q_{1}q_{2}\jmath '_{l}(1)y'_{l}(2) +\frac{\imath\omega\xi}{R}
\left(\frac{\rho _{1}\rho _{2}}{\rho _{1}-\rho _{2}}\right) ^{2}
\left(\frac{q_1}{\rho _1}\jmath '_{l}(1)y_{l}(2) - 
\frac{q_2}{\rho _1}\jmath _{l}(1)y'_{l}(2)\right)\right\} 
\label{equ24} 
\end{eqnarray}
The argument in $\jmath _l$ and $y_l$ is either $q_{1}R$ or $q_{2}R$. 
\par Since the general expressions are rather cumbersome, we concentrate on 
the wavelength limit $q_{1,2}R\ll 1$. In this limit the main contribution into 
the total scattering amplitude comes from lowest harmonics, $l=0$ and $l=1$ 
if $\xi =0$ and from $l=0$ alone if $\xi =\infty$. The partial $l=0$ 
amplitude is given by 
\begin{equation}
f_{0}=-\frac{R}{3}\, (q_{2}R)^{2}\, \frac{
(1-\rho _{2}c_{2}^{2}/\rho _{1}c_{1}^{2})+3(1-\omega _{0}^{2}/
\omega ^{2})(Y_{2}\xi)^{2} -3\imath Y_{2}\xi /(q_{2}R) }
{1+(1-\omega _{0}^{2}/\omega ^{2})^{2}(\omega R\rho _{2}\xi)^{2} }
\label{equ25}
\end{equation}
where $\omega _0$ is the normal frequency of the radial pulsations determined 
by Eq. (\ref{equ9}). The total scattering cross-section $\sigma _{tot}^{(0)} = 
\sigma _{el}^{(0)} +\sigma _{in}^{(0)}$ consists of two terms, elastic one 
\begin{equation}
\sigma _{el}^{(0)}=4\pi R^{2}\left(\frac{\omega R}{3c_2}\right) ^{2}
\frac{(1-\rho _{2}c_{2}^{2}/\rho _{1}c_{1}^{2})^{2}(\omega R/c_{2})^{2} 
+9(Y_{2}\xi)^{2} }
{1+(1-\omega _{0}^{2}/\omega ^2)^{2}(\omega  R\rho _{2}\xi )^2}
\label{equ26}
\end{equation}
and the other inelastic one meaning an additional absorption of sound if 
$0<\xi\ <\infty$ 
\begin{equation}
\sigma _{in}^{(0)}=4\pi R^2\frac{Y_2\xi}
{1+(1-\omega _{0}^{2}/\omega ^{2})^{2}(\omega R\rho _{2}\xi )^2} 
\label{equ27}
\end{equation}
\par For $\xi =0$, Eqs. (\ref{equ25}--\ref{equ27}) reduce to the well-known 
case of two immiscible liquids \cite{2LL}. The involvement that $\xi\neq 0$ 
leads to an enhancement of sound scattering, making it temperature-dependent 
due to $\xi =\xi (T)$. At large $\xi$ satisfying 
$Y_{2}\xi\gg (q_{2}R)^{-1}\gg 1$ the cross-section tends to 
\begin{equation}
\sigma =4\pi R^2\omega ^{4}/(\omega ^{2}-\omega _{0}^{2})^2 
\label{equ28}
\end{equation}
crossing over for $\omega\gg\omega _{0}$ into the cross-section corresponding 
to the scattering $\sigma =4\pi R^2$ on an impermeable sphere of radius $R$. 
The latter becomes clear if we take into account that the sound wave 
perturbation cannot penetrate into a drop for $\xi =\infty$. 
\par The corresponding contribution from $l=0$ into the variation of the sound 
velocity due to {\it c}-drops is given by
\begin{equation}
\frac{\Delta c_{2}^{(0)}}{c_2}= \frac{1}{2V}\sum _{R}\frac{4\pi R^3}{3}\cdot 
\frac{(1-\rho _{2}c_{2}^{2}/\rho _{1}c_{1}^{2}) 
+ 3(1-\omega _{0}^{2}/\omega ^{2})(Y_{2}\xi)^2}
{1+(1-\omega _{0}^{2}/\omega ^{2})^{2}(\omega R\rho _{2}\xi )^2 }
\label{equ29}
\end{equation}
As $Y_{2}\xi >\mid 1-\rho _{2}c_{2}^{2}/\rho _{1}c_{1}^{2}\mid\sim 1$, 
$\Delta c_{2}^{(0)}/c_{2}$ becomes temperature- and frequency-dependent. The 
strong effect gains when $Y_{2}\xi\sim c_{2}/\omega R\gg 1$, 
\begin{equation}
\Delta c_{2}^{(0)}/c_{2}\rightarrow (1/2V)\sum _{R} 4\pi R (c_{2}/\omega)^2 
\label{equ30}
\end{equation}
For $R=$ 10$\mu$m and $\omega =$ 6.3MHz, $c_{2}/\omega R\approx 3$. So we may 
expect an anomalous behavior in the scattering and sound velocity for a 
suspension of $\mu$m-sized {\it c}-drops at $T<$30mK. 
\par Concerning the contribution from harmonic $l=1$, the partial scattering 
amplitude $f_1$ is given by 
\begin{equation}
f_1 = -\frac{R}{3}\, (q_{2}R)^{2}\, \frac{
(\rho _{2}-\rho _{1})(2\rho _{1}+\rho _{2}) 
+(\omega R\xi\rho _{1}\rho _{2})^{2} 
-3\imath\omega R\xi\rho _{1}^{2}\rho _{2} }
{(2\rho _{1}+\rho _{2})^{2} + (\omega R\xi\rho _{1}\rho _{2})^2 }
\label{equ31}
\end{equation}
Accordingly, we have the elastic and inelastic contributions into the total 
cross-section $\sigma _{tot}^{(1)}=\sigma _{el}^{(1)}+\sigma _{in}^{(1)}$ 
\begin{eqnarray}
\sigma _{el}^{(1)} & = & \frac{4\pi R^2}{3}\left(\frac{\omega R}{c_2}\right) ^4 
\frac{(\rho _{2}-\rho _{1})^{2} + (\omega R\xi\rho _{1}\rho _{2})^2}
{(2\rho _{1}+\rho _{2})^{2} + (\omega R\xi\rho _{1}\rho _{2})^2} 
\nonumber \\
\sigma _{in}^{(1)} & = & 12\pi R^{2}\left(\frac{\omega R}{c_2}\right) ^{4} 
\frac{\omega R\xi\rho _{1}^{2}\rho _2 }
{(2\rho _{1}+\rho _{2})^{2} + (\omega R\xi\rho _{1}\rho _{2})^2}  
\end{eqnarray}
The variation of the sound velocity from $l=1$ equals 
\begin{equation}
\frac{\Delta c_{2}^{(1)}}{c_2} = \frac{3}{2V}\sum _{R}
\frac{4\pi R^3}{3}\cdot\frac{(\rho _{2}-\rho _{1})(2\rho _{1} +\rho _{2}) 
+ (\omega R\xi\rho _{1}\rho _{2})^2 }
{(2\rho _{1}+\rho _{2})^{2} + (\omega R\xi\rho _{1}\rho _{2})^2} 
\label{equ33} 
\end{equation}
Note only that in the $Y_{2}\xi\gg 1$ limit the contribution from harmonic 
$l=1$ is weaker than from $l=0$ and remains of the same order of the 
magnitude. As $\xi$ grows, $\sigma$ and $\Delta c_{2}/c_2$ increase. 
\par Finally, we consider higher harmonics with $l\geq 2$. The partial 
scattering amplitudes reflect the normal modes of the shape oscillations 
\begin{equation}
f_{l}= -\frac{R}{2l+1}\frac{1}{(2l-1)!!}\left(\frac{\omega R}{c_2}\right) ^{2l}
\frac{l(\rho _{2}-\rho _{1})-\imath\omega R\xi\rho _1}
{(l+1)\rho _{1} + l\rho _2}\cdot
\frac{\omega ^{2}+ \alpha l(l-1)(l+2) /(\rho _{2}-\rho _{1})R^3 }
{(\omega ^{2}-\omega _{0}^{2})^{2} -\imath\omega\tau _{l}
(\omega ^{2}-\omega _{\infty}^{2}) }   
\label{equ34}
\end{equation}
Here $\omega _0$, $\omega _{\infty}$, and $\tau _l$ are defined by 
Eqs. (\ref{equ13}--\ref{equ15}). Obviously, in spite of the resonance 
character 
the contribution of higher harmonics into $\sigma$ and $\Delta c_{2}/c_2$ 
is much smaller. However, some resonance  behavior as a function of $\omega$ 
may be observable for a homogeneous distribution over the sizes of drops.  
\newline
({\bf 4}) To conclude, we have calculated the oscillation spectrum of a 
liquid drop in a phase-se\-pa\-ra\-ted fluid with the highly mobile interface. 
It is found that the spherically symmetrical pulsations are strongly softened 
as compared with the case of two immiscible fluids. The sufficiently small 
drops can prove to be unstable. The nonspherical shape oscillations can be 
described as an oscillatory process with some effective relaxation time 
depending on the growth coefficient. The finite value of the growth 
coefficient leads both to the frequency shift and to an additional damping 
of the drop oscillations. 
\par In addition, we have found the cross-section of sound scattering 
on a drop and the variation of the sound velocity in a suspension of 
{\it c}-drops. Compared with the immiscible fluids, we should note that the 
total cross-section and, correspondingly, sound absorption enhance especially 
for large growth coefficients $Y_2\xi\geq 1$. A qualitative feature for 
the finite magnitudes of the growth coefficient is an appearance of the 
inelastic component of scattering. The sound velocity in a suspension of 
{\it c}-drops grows with an increase of the growth coefficient, displaying 
also a frequency- and temperature-dependent behavior. Thus we may expect 
anomalous behavior of a sound wave propagating across the two-phase system 
of $\mu$m-sized {\it c}-drops suspended in the {\it d}-phase for frequencies 
of 1--10MHz and larger at temperatures below about 30--50mK due to high 
mobility of the interface between the {\it c}- and {\it d}-phases. 
\par The work is supported by Yamada Foundation. 

\end{document}